\documentclass[12pt,aps,prd,superscriptaddress,showpacs,longbibliography,floatfix,nofootinbib]{revtex4-1}
\pdfoutput=1

\usepackage[utf8]{inputenc}
\usepackage{slashed}
\pdfoutput=1
\usepackage{hyperref}

\usepackage{color}
\usepackage{graphicx}   
\usepackage{bm}
\usepackage{amsmath}
\usepackage{amsfonts}
\usepackage{hyperref}

\def\be{\begin{equation}}
\def\ee{\end{equation}}
\def\ba{\begin{eqnarray}}
\def\ea{\end{eqnarray}}

\begin{document}

\title{Mass From Nothing\\
\normalsize{Part I: The Abelian Higgs Model}}

\author{Paul Romatschke,}
\author{Chun-Wei Su,}
\author{and Ryan Weller}
\affiliation{Department of Physics, University of Colorado, Boulder, Colorado 80309, USA}

\begin{abstract}
  We study the Abelian Higgs model with multiple scalar fields, but without mass terms. Solving the model non-perturbatively order-by-order in the number of scalar fields, we find that radiative corrections generate masses for the scalar and gauge boson, without spontaneous symmetry breaking. The mass scales are set by the $\Lambda$-parameter of the electroweak running coupling, thereby naturally avoiding the hierarchy problem. No part of our calculation employs a weak-coupling expansion, and we find that the perturbative vacuum is metastable, and hence must decay to the stable non-perturbative vacuum of the theory, which we identify. Although the field content of our Lagrangian is standard, our results predict the existence of two heavy scalar resonances in addition to the Higgs. We believe that these predicted resonances will ultimately allow experimentalists to  discriminate between our method and standard solutions of the Higgs model.
\end{abstract}
\maketitle

\section{Introduction}

In the Standard Model of Physics, the electroweak gauge bosons obtain their masses through the Higgs mechanism \cite{Englert:2014zpa}. The standard implementation of the Higgs mechanism in electroweak theory is through spontaneous symmetry breaking (SSB) \cite{Maas:2017wzi}, in which the Higgs field takes on a non-vanishing vacuum-expectation value:
\be
\langle \phi \rangle = {\rm VEV} \neq 0\,.
\ee

There are several well-documented \textbf{issues with the standard implementation of the Higgs mechanism} for generating gauge boson masses:
\begin{itemize}
  \item
  The symmetry broken by SSB would be a continuous symmetry (gauge symmetry), which is explicitly forbidden by \textbf{Elitzur's theorem} \cite{Elitzur:1975im}, as pointed out in detail in Ref.~\cite{Kondo:2018qus}
\item
  The Higgs VEV is engineered through a Higgs potential involving a ``\textbf{tachyonic}'' mass term:
  \be
  \label{Vphi}
  V(\phi)=-\mu^2 \phi^2+\lambda \phi^4\,.
  \ee
  No plausible reason for the sign of $\mu^2$ is given other than that ``it works''!
\item
  The mass of the Higgs boson (and its VEV) are set (by the choice of $\mu$) to the electroweak scale, which is much smaller than the Planck scale. This issue is known as the ``\textbf{hierarchy problem}'' \cite{Maas:2017wzi,Koren:2020pio}
\item
  The standard implementation proceeds via the weak-coupling (perturbative) expansion around the classical minimum of (\ref{Vphi}). Experimental data suggests that this minimum is \textbf{only metastable} with respect to decay to other, non-perturbative vacuua \cite{Buttazzo:2013uya}.
  \end{itemize}

\textbf{In this work, we propose a different mechanism for mass generation} that
\begin{itemize}
\item
  Does not involve a non-vanishing vacuum-expectation value:
  \be
  \langle \phi \rangle =0\,.
  \ee
\item
  Does not involve SSB because no symmetries are broken
\item
  Does not involve a tachyonic mass term in the potential
  \be
  \mu^2=0\,.
  \ee
\item
  has the Higgs mass scale emerge as $\Lambda$-parameter of the electromagnetic running-coupling trough dimensional transmutation, thereby naturally avoiding (as opposed to``solving'') the hierarchy problem
\item
  Uses a large N expansion around a non-perturbative vacuum of the theory rather than the usual weak-coupling expansion around the metastable classical vacuum
  \end{itemize}

Because our proposed mechanism does not have any dimensionful parameters in the theory Lagrangian, we refer to our mechanism as generating \textbf{mass from nothing}.

The individual components of our mechanism have been suggested in one form or another before, but to our knowledge they have never been combined in a single calculation from beginning to the end. For instance, mass generation without SSB, also known as symmetric mass generation, has been discussed in the context of scalar and fermionic theories \cite{Frohlich:1981yi,Kondo:2018qus,Tong:2021phe,Butt:2021koj}. An emergent Higgs mass scale from radiative corrections has been discussed in the context of spontaneous symmetry breaking and is known as ``Coleman-Weinberg-mechanism'', cf. Ref.~\cite{Coleman:1973jx}. Finally, large N expansions in the number of fields rather than a weak-coupling expansion has been employed in pure scalar field theories in Refs.~\cite{Abbott:1975bn,Linde:1976qh}, leading to the discovery of non-perturbative stable minima. 

The catalyst of bringing these separate developments together into one single mechanism was the realization that scalar quantum field theory could be \textit{non-trivial} in four dimensions \cite{Romatschke:2023sce} (see also \cite{Symanzik:1975rz,Parisi:1975im,Bender:1999ek}), because the condensed-matter-inspired proof of quantum triviality \cite{Aizenman:2019yuo} leaves an important loophole exploited by large N scalar field theory \cite{Romatschke:2023ogd}. This has led to a recent uptick in interest related to scalar and fermionic large N theories, broadly defined \cite{Cea:2022zgs,Romatschke:2022jqg,Romatschke:2022llf,Haba:2023jhp,Berges:2023rqa,Grable:2023paf,Lawrence:2023woz,Jepsen:2023pzm,Sinner:2023bdm,Maiezza:2023mvb,Weller:2023jhc,Flodgren:2023tri,Kamata:2023opn,Romatschke:2023fax,Kan:2023lah,Mavromatos:2024ozk,Bajec:2024jez}.

Importantly, if scalar field theory in four dimensions is a bona-fide interacting quantum field theory in the continuum, then non-perturbative solutions of this theory are possible. Putting the theory on a lattice is one potential approach \cite{Romatschke:2023fax,Lawrence:2022afv}, but complicated by the existence of a severe sign problem. For this reason, non-perturbative analytic techniques, such as an expansion in a large number of scalar fields $N\gg1$ currently seem most promising \cite{Romatschke:2023ztk}.

For this reason, here we consider the Abelian Higgs model where multiple scalar fields are covariantly coupled to a single U(1) gauge boson. Thus our theory Lagrangian is completely standard, but our solution technique is not. Instead of using effective theory and a weak-coupling expansion, systematically expanding in the number of scalar fields $N$ allows us to reliably study the non-perturbative sectors of the theory. In this work, our goal is \textit{not} a direct phenomenological application of our findings, even though we believe this could be a viable research direction in its own right. Instead, we choose to consider the Abelian Higgs model because it will allow to elucidate the similarities and differences with respect to the standard weak-coupling solution of the same theory with minimal complexity. In a follow-up work, we will present the generalization of our mechanism to non-Abelian Higgs and full electroweak theory, including phenomenological consequences.

Even though the field content of our theory is standard, and even though the symmetries are standard, owing to our non-perturbative solution techniques we do find features that are not present in the standard weak-coupling approach. In particular, it will be shown in this paper that in addition to the standard field content of the theory, the non-perturbative solution \textbf{predicts} the presence of \textbf{two scalar resonances} at masses of approximately $1.84$ and $2.63$ times the Higgs boson mass. We believe that these additional scalar particles are prime candidates for future experimental verification/falsification of our method.

Many experts trained in standard perturbative quantum field theory will be unfamiliar with the recent technical developments for many component scalar field theories. While we point out the most important new steps including references in this paper, a fully pedagogical treatment is beyond the scope of this work. For this reason, we suggest the interested reader browses the lecture notes on large N field theory \cite{Romatschke:2023ztk} for context, more detail, and additional references.

\section{Setup of the Abelian Higgs Model}

We choose to work in Euclidean field theory in order to give a non-perturbative definition of the full quantum field theory in terms of the path-integral representation of the partition function. Observables in Minkowski space are then obtained via the usual analytic continuation of the Euclidean quantities, cf.~\cite[chap. 8]{Laine:2016hma}. Let us now get more specific about the Abelian Higgs model, and first consider the Euclidean Lagrangian density of a complex scalar $\Phi$ covariantly coupled to an Abelian gauge field $A_\mu$:
\be
{\cal L}_E=\left(D_\mu \Phi\right)^\dagger\left(D_\mu \Phi\right)+\frac{1}{4}F_{\mu\nu}F_{\mu\nu}\,,\quad D_\mu\equiv \partial_\mu - i e_B A_\mu\,,
\ee
where $F_{\mu\nu}$ is the Abelian field-strength tensor and $e_B$ is the (bare) electromagnetic coupling parameter. The theory Lagrangian is invariant under the following local gauge transformation:
\be
\label{GF}
  \Phi(x)\rightarrow \Phi^\prime(x)=e^{i  \alpha(x) e_B}\Phi(x)\,,\quad A_\mu(x)\rightarrow A_\mu^\prime(x)=A_\mu(x)+\partial_\mu \alpha(x)\,.
  \ee
  Since $\Phi(x)$ is locally a complex number, we can use the above symmetry to  phase-rotate $\Phi(x)$ such that it is purely real, in complete analogy to what is performed in the standard Higgs mechanism implementation \cite{Langacker:2017uah}, see also \cite{vanEgmond:2023lnw}. This amounts to a gauge fixing procedure, but we perform the gauge fixing \textit{on the scalar degrees of freedom} rather than the gauge field $A_\mu$\footnote{We thank G.~Sundaresan for pointing out that this gauge fixing choice is described in Ref.~\cite{Faddeev:1980be}, chapter 1.1.}. This simplifies the following calculation, because the resulting Faddeev-Popov ghosts are not dynamic.

  We now want to introduce gauge fixing in the usual way, e.g. by inserting a $\delta$ function that imposes a condition on $\Phi(x)$ to fix the gauge, e.g. through the choice
  \be
  G[\Phi(x)]=0\,.
  \ee
  With the gauge transformation parametrized as in (\ref{GF}) by the function $\alpha$, this implies we insert
  \be
  1=\int {\cal D}\alpha \delta\left(G[\Phi(x)]\right)\left|{\rm det}\left(\frac{\delta G[\Phi(x)]}{\delta \alpha}\right)\right|\,,
  \ee
  into the path integral. To be specific, let us choose
  \be
  G[\Phi(x)]={\rm Im}\Phi(x)\,,
  \ee
  so that
  \be
  {\rm det}\left(\frac{\delta G[\Phi(x)]}{\delta \alpha}\right)=\prod_x e_B{\rm Re} \Phi(x)\,.
  \ee
  Writing $\Phi(x)=\phi(x)+i \phi_2(x)$ with $\phi(x),\phi_2(x)$ the real and imaginary parts of $\Phi(x)$, the partition function after gauge fixing thus becomes 
  \be
  \label{one1}
  Z=\int {\cal D}\phi(x) \left|\phi(x)\right| e^{-\int d^4x {\cal L}_{gf}}\,,
  \ee
  where we have absorbed the constant term $e_B$ into the path integral measure, and the gauge-fixed Euclidean Lagrangian density  is given by
  \be
  {\cal L}_{gf}=\left(D_\mu \phi\right)^\dagger\left(D_\mu \phi\right)+\frac{1}{4}F_{\mu\nu}F_{\mu\nu}\,,\quad D_\mu\equiv \partial_\mu - i e_B A_\mu\,,
  \ee
  where $\phi(x)$ is now a single \textit{real} scalar field. One can recognize the term $|\phi(x)|$ in (\ref{one1}) to correspond to the Faddeev-Popov determinant, and it is possible to exponentiate it using the usual introduction of Faddeev-Popov ghosts\footnote{Note that our determinant as written would be singular for all points where $\phi(x)=0$. However, the manifold on which $\phi(x)$ vanishes has lower dimension than our target dimension $d=4$. For this reason, the singular points in the determinant do not contribute to the partition function in the infinite volume limit of the theory. We thank Cliff Taubes for pointing this issue out to us.}. However, since the determinant is so simple, introduction of the ghosts is not necessary. In fact, in many discussions of the Higgs mechanism in the literature, the Faddeev-Popov determinant from fixing the scalar to be real is said to be ``absorbed in the measure of the path integral'', cf. the Lagrangian in the ``polar representation'' in Ref.~\cite{Langacker:2017uah}. Essentially, this is just a re-definition of the potential for the Higgs field. Because of this prevailing choice in the literature, and to be absolutely clear that this is not the source of mass generation in our setup, we adopt the same convention (packing the Faddeev-Popov determinant into a redefinition of the Higgs field potential) for the main part of our work. However, for those interested in its effect, we redo the calculation with the Faddeev-Popov determinant included in appendix \ref{app1}.

  It is now easy to generalize the above theory from one complex scalar field to a theory consisting of $N$ complex scalar fields. We consider $N$ complex scalar fields which are ``phase-locked'', e.g. can be represented as $\vec{\Phi}=\vec{\phi}e^{i e_B \alpha(x)}$ with the same complex phase for all $N$ fields. This is to ensure that the model has $N+1$ degrees of freedom rather than $2N$ degrees of freedom, and we can use the gauge fixing procedure as above.  After gauge-fixing, we end up with a theory involving N real scalar fields $\vec{\phi}=\left(\phi_1,\phi_2,\ldots\phi_N\right)$. Thus in the following we study the theory defined by the Euclidean action
  \be
  \label{def}
  S_E=\int dx \left[(D_\mu \vec{\phi})^\dagger D_\mu \vec{\phi}+ \frac{\lambda_B}{N} (\vec{\phi} \cdot \vec{\phi})^2+\frac{1}{4}F_{\mu\nu}F_{\mu\nu} \right]\,,
  \ee
  where in addition to the kinetic terms we have also included a potential term for the scalars with $\lambda_B$ the bare scalar self-coupling parameter. Note that no mass term is included in the potential, nor will it be included later on in the derivation, so there are no dimensionful parameters in the theory.
  
 The gauge covariant derivative and field strength tensor are defined by
  \be
  D_\mu=\partial_\mu -i \frac{e_B}{\sqrt{N}} A_\mu\,,\quad
  F_{\mu\nu}=\partial_\mu A_\nu-\partial_\nu A_\mu\,,
  \ee
  where $e_B$ is the gauge field coupling parameter. 
%
 %
%
%
  For the case $N>1$, the model is related to extensions of the Standard Model \cite{Workman:2022ynf}, but here N is taken as a book-keeping parameter in order to organize a non-perturbative solution of the model. The scaling of the couplings $\lambda_B,e_B$ with N has been chosen such that the theory possesses a well-defined limit for $N\rightarrow \infty$.

  It should be pointed out that -- usually -- path integrals for gauge theories suffer from flat directions. Because we have chosen to fix the gauge freedom as a condition on the scalar field rather than the vectors, the resulting path integral must be free of any flat directions. We show below that this is indeed the case.

\section{Calculation}
  
  Following Ref.~\cite{Romatschke:2023ztk}, it is useful to trade the quartic scalar coupling with respect to an additional auxiliary field. This can be done in terms of a mathematical identity:
  \be
  \label{HS}
  e^{-\frac{\lambda_B \left(\vec{\phi} \cdot \vec{\phi}\right)^2}{N}}=\int d\zeta e^{-i \zeta \vec{\phi}\cdot \vec{\phi}-\frac{N \zeta^2}{4\lambda_B}}\,.
  \ee
  After using (\ref{HS}), the partition function of the theory becomes
  \be
\label{Zthree3}
  Z=\int {\cal D}\vec{\phi}{\cal D}A {\cal D}\zeta e^{-\int dx \left(\vec{\phi}\left[-\Box+i\zeta+\frac{e_B^2}{N}A_\mu A_\mu\right]\vec{\phi}+\frac{N \zeta^2}{4\lambda_B}+\frac{1}{4} F_{\mu\nu}F_{\mu\nu}\right)}\,.
  \ee

  In this form, the scalar part of the action is quadratic in $\vec{\phi}$ and can formally be integrated out:
  \be
  \label{flumS}
  Z=\int {\cal D}A {\cal D}\zeta e^{-\frac{N}{2}{\rm tr}\ln\left[-\Box+i\zeta+\frac{e_B^2}{N}A_\mu A_\mu\right]-\int dx \left(\frac{N \zeta^2}{4\lambda_B}+\frac{1}{4} F_{\mu\nu}F_{\mu\nu}\right)}\,.
  \ee

  So far, everything has been exact. However, in order to make progress, an approximation will have to be made.
  
\subsection{(Only) approximation: large N limit}

     In the leading large N limit, only terms proportional to $N$ in the action of (\ref{flumS}) contribute. This case is identical to the case of the pure scalar case treated a long time ago \cite{Linde:1976qh,Abbott:1975bn}, and recently revisited\cite{Romatschke:2022jqg,Romatschke:2022llf,Romatschke:2023sce,Weller:2023jhc}. Expanding the action systematically in powers of $\frac{1}{N}$, one finds in particular
 \be
     {\rm tr}\ln\left[-\Box+i\zeta +\frac{e_B^2}{N}A_\mu A_\mu\right]=
     {\rm tr}\ln\left[-\Box+i\zeta\right]+\frac{e_B^2}{N}{\rm tr}\left[\Delta(x,y) A_\mu(x) A_\mu(x)\right]+{\cal O}\left(\frac{1}{N^2}\right)\,,
     \ee
     where
     \be
     \left[-\Box+i\zeta\right]\Delta(x,y)=\delta(x-y)\,.
     \ee
  
     We are particularly interested in studying mass generation for the U(1) gauge field $A_\mu$, so it is a nice feature of the large N expansion that the quadratic term in the gauge field in the action is automatically included at the NLO large N expansion level. Higher order contributions, while interesting when studying decay channels, are not of primary interest here, but will need to be included when calculating the width of the auxiliary scalar resonances (see discussion below).

     Truncating the action at (including) order ${\cal O}\left(N^0\right)$, one finds for the partition function
     \be
     Z_2=\int{\cal D}A {\cal D}\zeta e^{-S_2}\,,
     \ee
     with
     \ba
     \label{S2}
     S_2&=&\frac{N}{2}{\rm tr}\ln\left[-\Box+i\zeta\right]+\frac{e_B^2}{2}{\rm tr}\left[\Delta A_\mu A_\mu\right]+\int dx \left[\frac{N \zeta^2}{4\lambda_B}+\frac{1}{4} F_{\mu\nu}F_{\mu\nu}\right]\,.    
     \ea
     
      Expanding the auxiliary field as a global zero mode plus fluctuations gives
     \be
     \zeta(x)=\zeta_0+\xi(x)\,,\quad \int dx \xi(x)=0\,,
     \ee
     so that
     \be
     \frac{N}{2}{\rm tr}\ln\left[-\Box+i\zeta\right]+\int dx \frac{N \zeta^2}{4\lambda_B}=\frac{N}{2}\left[{\rm tr}\ln\left[-\Box+i\zeta_0\right]+\int dx \frac{\zeta_0^2}{2\lambda_B}
     +\int_{x,y} \xi D^{-1}\xi+\ldots\right]\,,
     \ee
     where the auxiliary field propagator $D(x-y)$ in Fourier space fulfills 
     \be
     \label{auxprop}
     D(k)=\frac{1}{\frac{1}{2\lambda_B}+\Pi(k)}\,,\quad
     \Pi(x)=\frac{1}{2}\Delta^2(x)\,,
     \ee
     with the scalar field propagator $\Delta=\left[-\Box+i \zeta_0+i \xi\right]^{-1}$ defined above. Now note that one can scale the fluctuations as
     \be
     \xi\rightarrow \frac{\xi}{\sqrt{N}}\,,
     \ee
     so that it becomes clear that the contribution from the fluctuations are suppressed as $N\rightarrow \infty$. This greatly simplifies the calculation. It implies that up to (including) next-to-leading order in a large N expansion, it is sufficient to replace the scalar propagator by its large N limit,
     \be
     \Delta(x)\rightarrow \Delta_0(x)\equiv \int \frac{d^4p}{(2\pi)^4}
     \frac{e^{i p \cdot x}}{p^2+ i\zeta_0}\,,
     \ee
     with $i\zeta_0$ taking the role of a mass (squared).

     The corresponding action, complete to next-to-leading order in large N, takes the form
     \ba
     \label{S3}
     S_3&=&\frac{N}{2}{\rm tr}\ln\left[-\Box+i\zeta_0\right]+\int dx \frac{N \zeta_0^2}{4\lambda_B}+\frac{e_B^2}{2}\Delta_0(x=0)\int dx  A_\mu(x) A_\mu(x)\\
     &&+\int dx \frac{1}{4} F_{\mu\nu}F_{\mu\nu}+\int dx dy \frac{\xi(x)D^{-1}(x-y)\xi(y)}{2}+{\cal O}(N^{-1})\,,     \nonumber
     \ea
     where now in addition to all the explicit N dependencies also the implicit subleading terms in large N in the auxiliary field have been expanded out.

     To this order in the large N expansion, the polarization tensors take the form
     \be
     \Pi(x)=\frac{1}{2}\Delta_0^2(x)\,,
     \ee
     and the partition function becomes
     \be
     Z=\int d\zeta_0 e^{-N {\cal V}[i\zeta_0]}\,,\quad e^{-N {\cal V}}=\int {\cal D}A  {\cal D}\xi e^{-S_3}\,.
     \ee
     In the large N limit, the partition function is given exactly in terms of the saddle point condition ${\cal V}^\prime[i\zeta_0]=0$. To leading order in large N, this saddle point condition is given by the derivative of $S_3$ wrt. $i\zeta_0$, or
     \be
     \frac{i\zeta_0}{2\lambda_B}=\frac{1}{2}\Delta_0(x=0)+{\cal O}(N^{-1})\,.
     \ee
     This result can  be used to simplify the expression for $S_3$, making the mass term for the gauge boson explicit:
     \ba
     S_3&=&\frac{N}{2}{\rm tr}\ln\left[-\Box+i\zeta_0\right]+\int dx \frac{N \zeta_0^2}{4\lambda_B}+\frac{e_B^2}{2\lambda_B} i\zeta_0 \int dx  A_\mu(x) A_\mu(x)\\
     &&+\int dx \frac{1}{4} F_{\mu\nu}F_{\mu\nu}+\int dx dy \frac{\xi(x)D^{-1}(x-y)\xi(y)}{2}+{\cal O}(N^{-1})\,.     \nonumber
     \ea

     In this form, the action is only quadratic in all the remaining fields ($A_\mu,\xi$) so one can immediately find the effective potential of the theory in terms of $\zeta_0$:
     \ba
        {\cal V}&=&\frac{1}{2}{\rm tr}\ln\left[-\Box+i\zeta_0\right]+\int dx \frac{\zeta_0^2}{4\lambda_B}+\frac{1}{2N}{\rm tr}\ln D^{-1}\nonumber\\
        &&+\frac{1}{2N}{\rm tr}\ln {\rm det}\left[\left(-\Box+ i\zeta_0 \frac{e_B^2}{\lambda_B}\right) \delta_{\mu\nu}+\partial_\mu \partial_\nu\right]+{\cal O}(N^{-2})\,,
        \ea
        where the determinant in the last term is over the Lorentz structure. In Fourier space representation it can be written as
     \ba
     \label{effpot}
        \frac{{\cal V}}{{\rm vol}}&=&\frac{1}{2}\int \frac{d^4p}{(2\pi)^4}\ln\left[p^2+i\zeta_0\right]+\frac{\zeta_0^2}{4\lambda_B}+\frac{1}{2N}\int \frac{d^4p}{(2\pi)^4}\ln \left[\frac{1}{2\lambda_B}+\Pi(p)\right]\nonumber\\
        &&+\frac{1}{2N}\int \frac{d^4p}{(2\pi)^4}\ln {\rm det}\left[\left(p^2+  i\zeta_0 \frac{e_B^2}{\lambda_B}\right) \delta_{\mu\nu}-p_\mu p_\nu\right]+{\cal O}(N^{-2})\,,
        \ea
        where ${\rm vol}$ indicates the space-time volume.
        We will use dimensional regularization to evaluate the remaining integrals in the following.

        \subsection{Evaluation of effective potential to NLO in large N}

        In dimensional regularization, one shifts the number of space-time dimension to be non-integer:
        \be
        d=4-2\varepsilon, \quad 0<{\varepsilon}\ll 1\,,
        \ee
        and takes the limit $\varepsilon\rightarrow 0$ at the end of the calculation.
        One hallmark of dimensional regularization is that it is only sensitive to logarithmic divergencies, which we will be using in the following. The heavy-lifting is performed by the analytic continuation of the Gamma function, which for clarity we explain through the following example. Consider the momentum integral
        \be
        \label{oneM}
        \mu^{2\varepsilon}\int \frac{d^d p}{(2\pi)^d}\frac{1}{p^2+m^2}=\frac{m^2 \Gamma\left(-1+\varepsilon\right)}{(4\pi)^{2}}  \left(\frac{m^2}{4\pi\mu^2}\right)^{-\varepsilon}
        \ee
        in $d=4-2\varepsilon$ dimensions, which has been recognized to correspond to an integral representation of the $\Gamma$ function. Here $\mu$ is a fictitious energy scale that was introduce such that the above integral has mass dimension $2$ even if $d=4-2\varepsilon$ is non-integer. An immediate consequence of dimensional regularization is that the above integral vanishes whenever $m=0$, unlike what one would find in cut-off regularization. This behavior is well-documented and understood: dimensional regularization will only report logarithmic divergencies, cf. Ref.~\cite{Romatschke:2019nmo}.

        For the case at hand, it is useful to integrate the above integral wrt to $m^2$ to find
        \be
        \mu^{2\varepsilon}\int \frac{d^d p}{(2\pi)^d}\ln (p^2+m^2)=\frac{m^4}{(4\pi)^{2}}\frac{\Gamma(-1+\varepsilon)}{2-\varepsilon} \left(\frac{m^2}{4\pi\mu^2}\right)^{-\varepsilon}\,,
        \label{dimmi}
        \ee
        which again vanishes for $m=0$.
        For the case $m\neq 0$, one can expand in powers of $\varepsilon$, finding
        \be
         \lim_{\varepsilon\rightarrow 0}\mu^{2\varepsilon}\int \frac{d^d p}{(2\pi)^d}\ln (p^2+m^2)=-\frac{m^4}{2 (4\pi)^{2}}\left(\frac{1}{\varepsilon}+\ln \frac{\bar\mu^2 e^{\frac{3}{2}}}{m^2}\right)\,,
        \label{dimmi2}
        \ee
        and where $\bar\mu^2=4\pi \mu^2e^{-\gamma_E}$ is the renormalization scale in the $\overline{\rm MS}$ scheme. Using these results, and introducing the short-hand notation
        \be
        m^2=i \zeta_0\,,
        \ee
        the effective potential in dimensional regularization becomes
         \ba
     \label{effpo2t}
        \frac{{\cal V}}{{\rm vol}}&=&-\frac{m^4}{64\pi^{2}}\left(\frac{16\pi^2}{\lambda_B}+\frac{1}{\varepsilon}+\ln \frac{\bar\mu^2 e^{\frac{3}{2}}}{m^2}\right)+\frac{1}{2N}\int \frac{d^4p}{(2\pi)^4}\ln \left[\frac{1}{2\lambda_B}+\Pi(p)\right]\nonumber\\
        &&+\frac{1}{2N}\int \frac{d^4p}{(2\pi)^4}\ln {\rm det}\left[\left(p^2+ m^2 \frac{e_B^2}{\lambda_B}\right) \delta_{\mu\nu}-p_\mu p_\nu\right]+{\cal O}(N^{-2})\,,
        \ea
        and the polarization tensor is given by
        \ba
        \label{pires}
        \Pi(p)&=&\frac{\mu^{2\varepsilon}}{2}\int \frac{d^dk}{(2\pi)^d}\frac{1}{(p-k)^2+m^2}\frac{1}{k^2+m^2}=\frac{\mu^{2\varepsilon}\Gamma(\varepsilon)}{2(4\pi)^{\frac{d}{2}}}\int_0^1dx \left(m^2+x(1-x)p^2\right)^{-\varepsilon}\,,\nonumber\\
        &=&\frac{1}{32\pi^2}\left(\frac{1}{\varepsilon}+\ln \frac{\bar\mu^2e^2}{m^2}-2 \sqrt{\frac{p^2+4m^2}{p^2}}{\rm atanh}\sqrt{\frac{p^2}{p^2+4m^2}}+\varepsilon \Pi_2(p)\right)\,,
        \ea
        where
        \ba
        \Pi_2(p)&=&\frac{1}{12}\left[\pi^2+6\ln^2\frac{\bar\mu^2}{m^2}+12 \ln \frac{\bar\mu^2}{m^2}\left(2-2 \sqrt{\frac{p^2+4m^2}{p^2}}{\rm atanh}\sqrt{\frac{p^2}{p^2+4m^2}}\right)\right.\nonumber\\
          &&\left.+6\int_0^1 dx \ln^2\left(1+x (1-x)\frac{p^2}{m^2}\right)\right]\,,
        \ea
        contains logarithms and polylogarithms with arguments depending on $\frac{p^2}{m^2}$.

        For the contributions of the gauge boson, it is easiest to introduce the orthogonal projectors \cite{Laine:2016hma,Kraemmer:2003gd}
        \be
        P_{\mu\nu}^T=\delta_{\mu\nu}-\frac{p_\mu p_\nu}{p^2}\,,\quad P_{\mu\nu}^L=\frac{p_\mu p_\nu}{p^2}\,,
        \ee
        in terms of which the determinant is straightforward to calculate, and one finds
        \be
        \ln {\rm det}\left[\left(p^2+ m_A^2 \right) \delta_{\mu\nu}-p_\mu p_\nu\right]=(d-1)\ln\left[p^2+m_A^2\right]+\ln \left[m_A^2 \right]\,,
        \ee
        where
        \be
        \label{madef}
        m_A^2= m^2 \frac{e_B^2}{\lambda_B}\,.
        \ee
        Since $\ln [m_A^2]$ is independent from $p^2$, the corresponding momentum integral vanishes in dimensional regularization, and we get for the effective potential
        \ba
     \label{effpo3t}
        \frac{{\cal V}}{{\rm vol}}&=&-\frac{m^4}{64\pi^2}\left(\frac{16\pi^2}{\lambda_B}+\frac{1}{\varepsilon}+\ln \frac{\bar\mu^2 e^{\frac{3}{2}}}{m^2}\right)+\frac{1}{2N}\int \frac{d^4p}{(2\pi)^4}\ln \left[\frac{1}{2\lambda_B}+\Pi(p)\right]\nonumber\\
        &&-\frac{3}{N}\frac{m_A^4}{64\pi^{2}}\left(\frac{1}{\varepsilon}+\ln \frac{\bar\mu^2 e^{\frac{5}{6}}}{m_A^2}\right)+{\cal O}(N^{-2})\,.
        \ea
        
        \subsection{Non-perturbative renormalization}

        The theory can be non-perturbatively renormalized as follows: first note that in the large N limit, only the first term in (\ref{effpo3t}) contributes. Hence, the large N renormalization condition in $\overline{\rm MS}$ must be given by
        \be
        \frac{1}{\lambda_B}=\frac{1}{\lambda_R(\bar\mu)}-\frac{1}{16\pi^2\varepsilon}+{\cal O}(N^{-1})\,,
        \ee
        so that the large N exact running coupling $\lambda_R(\bar\mu)$ is given by
        \be
        \label{LOrunco}
        \lambda_R(\bar\mu)=\frac{16\pi^2}{\ln \frac{\Lambda_{\overline{\rm MS}}^2}{\bar\mu^2}}+{\cal O}(N^{-1})\,.
        \ee
        The LO large N renormalization and running coupling are sufficient for calculating the contributions at ${\cal O}(N^{-1})$ in (\ref{effpo3t}). For instance, we find that -- using the above formulas for the bare and running coupling -- the combination
        \be
        \label{all1}
        \frac{1}{2\lambda_B}+\Pi(p)=\frac{1}{32\pi^2}\left(\ln \frac{\bar\Lambda_{\overline{\rm MS}}^2e^2}{m^2}-2 \sqrt{\frac{p^2+4m^2}{p^2}}{\rm atanh}\sqrt{\frac{p^2}{p^2+4m^2}}+\varepsilon \Pi_2(p)\right)
        \ee
        is automatically finite, and we may proceed to evaluate its momentum integral.

        For the renormalization to proceed, we first need to isolate the divergent part of the integral over the logarithm in (\ref{effpo3t}). This is not completely trivial in dimensional regularization, but has been done in Ref.~\cite{Romatschke:2024yhx}. In this work, we are also interested in evaluating the finite part of this integral, which is why we need to include the term proportional to $\varepsilon$ in (\ref{all1}). Following the steps outlined in appendix A of Ref.~\cite{Romatschke:2024yhx}, we find 
        \ba
        \frac{1}{2N}\int \frac{d^{d}p}{(2\pi)^d}\ln \left[\frac{1}{2\lambda_B}+\Pi(p)\right]&=&\frac{m^4}{32\pi^2 N}\left[-\frac{4}{\varepsilon}+\left(9+6\ln\frac{\Lambda_{\overline{\rm MS}}^2e^1}{m^2}\right)\ln \varepsilon\right.\nonumber\\
        &&\left.-4 \ln \frac{\bar\mu^2 e^{1}}{m^2}+f\left(\frac{\Lambda_{\overline{\rm MS}}^2e^{1}}{m^2}\right)\right]\,,
        \ea
        where $f\left(\frac{\Lambda_{\overline{\rm MS}}^2e^{1}}{m^2}\right)$ can be evaluated numerically for any $m^2$, e.g. $f(1)\simeq8.42\ldots$.

        The effective potential thus reads
         \ba
     \label{effpo4t}
     \frac{{\cal V}}{{\rm vol}}&=&-\frac{m^4}{64\pi^2}\left[\frac{16\pi^2}{\lambda_B}+\frac{1}{\varepsilon}\left(1+\frac{8}{N}+\frac{3}{N}\frac{e_B^4}{\lambda_B^2}\right)-\frac{18}{N}\ln \varepsilon
     +\left(1+\frac{8}{N}\right)\ln \frac{\bar\mu^2 e^{\frac{3}{2}}}{m^2}+\frac{3}{N}\frac{e_B^4}{\lambda_B^2}\ln \frac{\bar\mu^2 e^{\frac{5}{6}}}{m_A^2}\right)\nonumber\\
        &&+\frac{6m^4}{32\pi^2 N}\ln \frac{\Lambda_{\overline{\rm MS}}^2 e^1}{m^2}\ln \varepsilon+\frac{m^4}{32\pi^2 N}\left(2+f\left(\frac{\Lambda_{\overline{\rm MS}}^2e^{1}}{m^2}\right)\right]+{\cal O}(N^{-2})\,.
     \ea
     Following the discussion on renormalizing the pure scalar O(N) model to NLO in large N in Ref.~\cite{Romatschke:2024yhx}, we write
     \be
     \label{rescalingM}
     m^2= m_H^2\left(1+\frac{6 \ln \varepsilon}{N}\right)\,,
     \ee
     to cancel the term proportional to $\ln m^2 \ln \varepsilon$ in the effective potential. In terms of $m_H^2$, the effective potential takes the form
      \ba
     \label{effpo5t}
     \frac{{\cal V}}{{\rm vol}}&=&-\frac{m_H^4}{64\pi^2}\left[\frac{16\pi^2}{\lambda_B}+\frac{1}{\varepsilon}\left(1+\frac{8}{N}+\frac{3}{N}\frac{e_B^4}{\lambda_B^2}\right)-\frac{18}{N}\ln \varepsilon
     +\left(1+\frac{8}{N}\right)\ln \frac{\bar\mu^2 e^{\frac{3}{2}}}{m_H^2}+\frac{3}{N}\frac{e_B^4}{\lambda_B^2}\ln \frac{\bar\mu^2 e^{\frac{5}{6}}}{m_A^2}\right)\nonumber\\
        &&+\frac{m_H^4}{32\pi^2 N}\left(2+f\left(\frac{\Lambda_{\overline{\rm MS}}^2e^{1}}{m_H^2}\right)\right]+{\cal O}(N^{-2})\,.
     \ea
     Non-perturbative renormalization then implies
     \be
     \label{NPdef}
        \frac{e_B^2}{\lambda_B}=\frac{e_R^2(\bar\mu)}{\lambda_R(\bar\mu)}={\rm const}=u^2\,,\quad \frac{1}{\lambda_B}=\frac{1}{\lambda_R(\bar\mu)}-\frac{1}{16\pi^2 \varepsilon}\left(1+\frac{8}{N}+\frac{3 u^4}{N}\right)+\frac{9}{8\pi^2 N}\ln \varepsilon\,,
        \ee
        and the running couplings are given by
        \be
        \label{runni}
        \lambda_R(\bar\mu)=\frac{16\pi^2}{\left(1+\frac{8}{N}+\frac{3 u^4}{N}\right)\ln \frac{\Lambda_{\overline{\rm MS}}^2}{\bar\mu^2}}+{\cal O}(N^{-2})\,,\quad
        \alpha(\bar\mu)=\frac{4\pi u^2}{\left(1+\frac{8}{N}+\frac{3 u^4}{N}\right)\ln \frac{\Lambda_{\overline{\rm MS}}^2}{\bar\mu^2}}+{\cal O}(N^{-2})\,,
        \ee
        where we have introduced the fine structure parameter $\alpha\equiv \frac{e^2}{4\pi}$. Up to (including) order $\frac{1}{N}$, the renormalized effective potential now is 
        \ba
     \label{effpo6t}
     \frac{{\cal V}}{{\rm vol}}&=&-\frac{m_H^4}{64\pi^2}\left[
       \left(1+\frac{8+3 u^4}{N}\right)\ln \frac{\Lambda_{\overline{\rm MS}}^2 e^{\frac{3}{2}}}{m_H^2}-\frac{3 u^4\ln u^2+2u^4+4+2 f\left(\frac{\Lambda_{\overline{\rm MS}}^2e^{1}}{m_H^2}\right)}{N}\right]\,.
     %
     \ea
     The value of $m_H^2$ is determined by the saddle point of ${\cal V}$. One finds two possible solutions:
     \be
     \label{saddles}
     m_H^2=0\,,\quad m_H^2=\exp{\left[1-\frac{2 (10.42+u^4+3 u^4 \ln u)}{N}\right]}\Lambda_{\overline{\rm MS}}^2\equiv \bar m_{H}^2\,,
     \ee
     where we have inserted the leading large N solution for $m_H^2$ to evaluate the argument of the function f. The first solution $m_H^2=0$ corresponds to the perturbative saddle of the theory, whereas the second saddle is non-perturbative. To decide which saddle is realized, one has to compare the effective potential evaluated at the saddles. One finds
     \be
     \label{nonPV}
     \frac{{\cal V}(m_H=\bar m_{H})}{\rm vol}=-\frac{1+\frac{8+3 u^4}{N}}{128\pi^2}\exp{\left[2-\frac{4\left(10.42+u^4+3 u^4 \ln u\right)}{N}\right]}\,,
     \ee
     whereas $\frac{{\cal V}(m_H=0)}{\rm vol}=0$.

     A crucial point is that the symmetry broken phase with $\langle \vec{\phi}\rangle \neq 0$ also has ${\cal V}=0$, which can be shown by direct calculation by letting $\vec{\phi}=\vec{\phi}_0+{\rm fluctuations}$ in (\ref{Zthree3}), cf. Refs.~\cite{Abbott:1975bn,inprep}. One finds that while the corresponding SSB saddle with $\langle \vec{\phi}\rangle \neq 0$ exists, it requires $m_H=0$, and is therefore identical to the perturbative saddle in our calculation.

     Comparing the effective potential, once evaluated at the perturbative saddle ${\cal V}=0$, and once evaluated at the non-perturbative saddle (\ref{nonPV}), one finds that for all N, the non-perturbative saddle has the lower free energy. As a consequence, we find that at least in the large N limit, the non-perturbative vacuum is the preferred phase, while the perturbative vacuum is at best meta-stable. While our result only strictly holds in the large N limit of the Abelian Higgs model, it is curious to note that our stability analysis agrees with the expectation that the perturbative Higgs vacuum in the full Standard Model is not stable, cf. Ref.~\cite{Buttazzo:2013uya}.

     \subsection{Vector boson and Higgs masses}

     The non-vanishing value of $\bar m_H$ implies a non-zero mass for the scalar $\vec{\phi}$ at leading order in large N:
     \be
     \label{LOhiggs}
     \bar m_{\rm Higgs}^2=\Lambda_{\overline{\rm MS}}^2 e^1+{\cal O}(N^{-1})\neq 0\,.
     \ee
     It is possible to evaluate the $\frac{1}{N}$ corrections to the mass of the scalars by evaluating the pole mass to order (including) ${\cal O}(N^{-1})$. This is most easily accomplished by consistently incorporating $\frac{1}{N}$ on the level of the action, via the R2-level resummation scheme \cite{Romatschke:2019rjk,Romatschke:2019ybu,Romatschke:2021imm}, which we discuss in the next section.

     The non-perturbative saddle found in the preceding subsection also implies a non-vanishing expectation value for the vector boson mass. Specifically,
     \be
     \label{LOma}
     \bar m_A^2\equiv \bar m_{\rm Higgs}^2 u^2+{\cal O}(N^{-1})\neq 0\,.
     \ee

Both the Higgs and gauge boson mass arise without any spontaneous symmetry breaking, which is anyhow forbidden by Elitzur's theorem \cite{Elitzur:1975im}.
     
\section{Including 1/N corrections}

Including 1/N corrections systematically necessarily requires infinite-order resummations of perturbative Feynman diagrams. At the NLO large N level, this can be implemented straightforwardly by starting from (\ref{Zthree3}) with $\zeta(x)=\zeta_0+\xi(x)$ and adding and subtracting self-energies for the three types of field $\vec{\phi},\xi,A_\mu$:
\ba
\label{R2def}
S_E&=&S_{R2,0}+S_{R2,I}\\
S_{R2,0}&=&\frac{N \zeta_0^2}{4\lambda_B}\times {\rm vol}+\int dx \left(\frac{1}{2}\vec{\phi}\Delta_{R2}^{-1}\vec{\phi}+\frac{N}{2}\xi D^{-1}\xi+
\frac{1}{2}A_\mu G_{\mu\nu}^{-1} A_\nu\right)\,,\nonumber\\
S_{R2,I}&=&-\int dx dy \left(\frac{1}{2}\vec{\phi}(x)\Sigma(x-y)\vec{\phi}(y)+\frac{N}{2}\xi(x)\Pi(x-y)\xi(y)+\frac{1}{2}A_\mu(x)\Pi_{\mu\nu}(x-y)A_\nu(y)\right)\nonumber\\
&&+\int dx \left(\frac{e_B^2}{2N}\vec{\phi}^2 A_\mu A_\mu+\frac{i}{2} \xi \vec{\phi}^2\right)\,,
\ea
where we rescaled $\vec{\phi}\rightarrow \frac{\vec{\phi}}{\sqrt{2}}$ to have a canonical normalization of the scalar fields.
Here the propagators for the scalars, auxiliary and gauge boson in Fourier space are given by
\be
\Delta_{R2}(k)=\frac{1}{k^2+i\zeta_0+\Sigma(k)}\,,\quad
D(k)=\frac{1}{\frac{1}{2\lambda_B}+\Pi(k)}\,,\quad
G^{-1}_{\mu\nu}(k)=k^2\delta_{\mu\nu}-k_\mu k_\nu+\Pi_{\mu\nu}(k)\,.
\ee
Note that the auxiliary field propagator $D(k)$ was already encountered before in (\ref{auxprop}). The self-energy corrections themselves are evaluated to leading order in the large N expansion, finding
\be
\Sigma(x)=\frac{e_B^2}{N}\delta(x) G_{\mu\mu}(x)+ \frac{1}{N}D(x)\Delta_{R2}(x)\,,\quad
\Pi_{\mu\nu}(x)=\delta_{\mu\nu} e_B^2\delta(x)\Delta_{R2}(x)\,.
\ee
along with the result for $\Pi(k)$ already found in (\ref{pires}). Note that since $\Sigma(x)\propto \frac{1}{N}$, the self-energy contribution in $\Delta_{R2}$ can be neglected for the purpose of calculating $\Sigma,\Pi_{\mu\nu}$, so that to this order in the large N expansion
\be
\Sigma(x)=\frac{e_B^2}{N}\delta(x)G_{\mu\mu}(x)+ \frac{1}{N}D(x)\Delta_{0}(x)\,,\quad
\Pi_{\mu\nu}(x)= m_A^2\delta(x)\delta_{\mu\nu}\,,
\ee
where $m_A^2$ was defined in (\ref{madef}). Note that this implies  in dimensional regularization
\be
G_{\mu\nu}(x)=\mu^{2\varepsilon}\int \frac{d^dp}{(2\pi)^d}\frac{e^{i p x}}{p^2+m_A^2}P_{\mu\nu}^T(p)\,,
\ee
so that
\be
\Sigma(x)=\frac{e_B^2}{N}(d-1)\delta(x)\mu^{2\varepsilon}\int\frac{d^dp}{(2\pi)^d}\frac{1}{p^2+m_A^2}+\frac{1}{N}D(x)\Delta_0(x)\,.
\ee
The first term is just a constant and may be calculated using (\ref{oneM}) as 
\be
e_B^2\int\frac{d^dp}{(2\pi)^d}\frac{1}{p^2+m_A^2}=-\frac{e_B^2}{\lambda_B} \frac{m_A^2\lambda_B}{16\pi^2\varepsilon}+{\cal O}(e_B^2\varepsilon^0)=u^2m_A^2+{\cal O}(\varepsilon)\,,
\ee
where we used non-perturbative renormalization (\ref{NPdef}). As a consequence we have in Fourier space
\be
\label{sf}
\Sigma(k)=\frac{1}{N}\left(3 u^2 m_A^2+\int \frac{d^4p}{(2\pi)^4}D(p)\Delta_0(p-k)\right).
\ee
From $\Delta_{R2}$ above, it is clear that the pole mass $m_{\rm Higgs}$ for the scalars $\vec{\phi}$ has to fulfill
\be
m_{\rm NLO}^2=m^2+\Sigma_R\,,
\ee
with $m^2=i\zeta_0$ the saddle point location and $\Sigma_{R}$ the retarded self-energy evaluated at the pole mass. Extracting the zero-momentum contribution from $\Sigma$ using (\ref{sf}) leads to
\be
m_{\rm NLO}^2=m^2+\Sigma(k=0)+\Delta \Sigma_R\,,
\ee
where $\Delta \Sigma(k)=\Sigma(k)-\Sigma(0)$. After the rescaling (\ref{rescalingM}), this becomes
\be
\label{polemass}
m_{\rm NLO}^2=m_H^2+m_H^2\frac{6}{N}\ln \varepsilon+\Sigma(k=0)+\Delta\Sigma_R\,,
\ee
which includes all corrections to (including) ${\cal O}(N^{-1})$. We recall that $\Sigma(k=0)$ contains a divergent contribution $-\frac{6}{N}m_H^2\ln \varepsilon$ that precisely cancels the divergent term introduced from (\ref{rescalingM}), so that the $\frac{1}{N}$ contribution to the pole mass is finite \cite{Romatschke:2024yhx}. It can be calculated by using the spectral representation of the propagator, and we will report on this result in a future work.

\subsection{Emergent resonances}
\label{sec:resonances}

There are emergent propagating particles appearing at the R2 resummation level that are not obvious from the (\ref{def}). To see this, consider the propagator for the auxiliary scalar $\xi$, given above. After non-perturbative renormalization, evaluating (\ref{all1}) for $m^2=\bar m_{\rm Higgs}^2=\Lambda_{\overline{\rm MS}}^2 e^1+{\cal O}(\frac{1}{N})$ one has
\be
D^{-1}(p)=\frac{1}{32\pi^2}\left(1-2 \sqrt{\frac{p^2+4\bar m_{\rm Higgs}^2}{p^2}}{\rm atanh}\sqrt{\frac{p^2}{p^2+4\bar m_{\rm Higgs}^2}}\right)\,.
\ee
Particle degrees of freedom can be identifies from the zeros of $D^{-1}$ when analytically continuing the momenta from Euclidean to real frequencies $\omega$, cf.~\cite{Laine:2016hma,Romatschke:2021imm}
\be
p^2\rightarrow (i \omega-0^+)^2=-\omega^2-i \omega 0^+\,.
\ee
Recovering the results found in the 1970s \cite{Abbott:1975bn}, one finds that the inverse auxiliary propagator has a zero for
\be
\omega\simeq 1.84 \bar m_{\rm Higgs}\,,
\ee
corresponding to an O(N) singlet or a bound state of $\vec{\phi}$ with mass (cf Ref.~\cite{Romatschke:2023sce})
\be
\bar m_{\rm bound}\simeq 3.03 \Lambda_{\overline{\rm MS}}\,.
\ee
This bound state is expected to acquire a non-vanishing width at ${\cal O}(N^{-1})$ in the Abelian Higgs model because decays into the massive vector bosons are allowed. For this reason, even though it is unconditionally stable to leading order in large N, we refer to this bound state as a resonance from now on.

In addition to this resonance, the real part of $D^{-1}(p^2=(i \omega-0^+)^2)$ vanishes for
\be
\omega\simeq 2.63 \bar m_{\rm Higgs}\,,
\ee
but the imaginary part of $D^{-1}$ does not. Taylor-expanding $D^{-1}$ near this frequency, we can match it to the form of a Breit-Wigner function, and we are led to identify
\be
\left|\frac{{\rm Re}D^{-1}(\omega)}{{\rm Im}D^{-1}(\omega)}\right|^2=\frac{(\omega-\bar m_{\rm resonance})^2}{\bar \Gamma_{\rm resonance}^2}\,,
\ee
indicating a broad resonance with a mass and width of
\be
\bar m_{\rm resonance}\simeq 4.3 \Lambda_{\overline{\rm MS}}\,,\quad
\bar \Gamma_{\rm resonance}\simeq 1.57 \bar m_{\rm Higgs}\simeq 2.59 \Lambda_{\overline{\rm MS}}\,.
\ee

Since the presence of these resonances is not obvious from the Lagrangian of the Abelian Higgs model, and since there is no mention of these resonances in the standard perturbative solution of this well-studied model, we expect that they could provide a possible experimental verification/falsification of our solution method. We do want to point out, however, that arguments in favor of an additional Higgs resonances have been given in Refs.~\cite{Cea:2022zgs,Consoli:2023hnw}.



        \section{Phenomenology and Conclusions}

        The Abelian Higgs model does not lend itself directly to particle physics phenomenology, but we nevertheless would like to study how numerical values emerge from our calculation. The only mass scale we use as an input is the electron mass $m_e\sim 0.511$ MeV, the rest are dimensionless couplings such as the fine structure constant $\alpha$, which we fix from (\ref{runni}) as 
 \be
        \alpha(\bar\mu=m_e)\simeq \frac{1}{137}=\frac{4\pi u^2}{\left(1+\frac{8}{N}+\frac{3 u^4}{N}\right)\ln \frac{\Lambda_{\overline{\rm MS}}^2}{m_e^2}}\,.
        \ee
        This leaves one more parameter undetermined. We choose to fix the ratio of coupling constants $u$ from (\ref{NPdef}) as
        \be
        u=0.362\,,
        \ee
        which is a constant of order unity. (Note that the phenomenological values we get are quite sensitive to the precise value of $u$, but this could be a feature of the model rather than a shortcoming.)

        With these choices, all model parameters are fixed. Using $N=1$, one finds that the electroweak scale is given by
        \be
        \Lambda_{\overline{\rm MS}}=132\ {\rm GeV}\,.
        \ee
        Furthermore, we have
        \be
        \label{pheno1}
\bar m_A=78.8\ {\rm GeV},\quad \bar m_{\rm Higgs}\simeq 217.7\ {\rm GeV}\,,\quad \bar m_{\rm bound}\simeq 400\ {\rm GeV}\,,
\ee
for the masses of the gauge boson, the Higgs boson and the predicted mass of the bound state. In addition, we predict a broad resonance with mass and width
\be
m_{\rm resonance}\simeq 572 {\rm GeV}\,,\quad \Gamma_{\rm resonance}\simeq 342{\rm GeV}\,.
\ee

While clearly not in agreement with the experimentally measured Higgs mass, these results are not qualitatively inconsistent with expectations. In particular, the partial NLO corrections from (\ref{saddles}) \textit{reduce} the value of the Higgs mass from its leading-order large N value.

There are many other choices of parameters that could lead to phenomenologically interesting results for the gauge boson and Higgs mass. The main result of our work, however, is that masses such as (\ref{pheno1}) arise naturally from the Abelian Higgs model \textit{in the absence of any mass parameter in the Lagrangian or symmetry breaking mechanism}. In particular, this generation of mass from nothing is a consequence of our non-perturbative solution of the model, and not due to any BSM ingredient. 

Because we solve the Abelian Higgs model in the continuum with the UV cut-off sent to infinity, our solution naturally avoids the hierarchy problem: both the Higgs and gauge boson mass are fixed by $\Lambda_{\overline{\rm MS}}$ times a number of order unity, despite (or rather because of) the consistent inclusion of radiative corrections. The key ingredient that allows this construction is the existence of quantum field theory with upside down potentials \cite{Ai:2022csx,Lawrence:2023woz,Romatschke:2023sce}. Defying classical intuition, this is a special case of so-called non-Hermitian field theories \cite{Bender:1998ke,Bender:2023cem}, which likewise defy classical intuition, but correctly describe experiment at least in lower dimensional systems, cf. Ref.~\cite{Rotter:2016,Xu:2016}.

Ultimately, since physics is a natural science, experiment will have to decide if our ``mass from nothing'' mechanism, or the perturbative Higgs mechanism with its extra parameters and UV-cutoff, is realized in nature. In this regard, the prediction of the resonances in our section \ref{sec:resonances} could play a key role, since no such extra particle degrees of freedom have been identified in the weak-coupling solutions of the Abelian Higgs model.

In order to make contact with experiment, more work on our mechanism is needed. For instance, we need to show that the ``mass from nothing'' mechanism generalizes to the case of the non-Abelian Higgs model, and ultimately to the full electroweak sector of the Standard Model of Physics. Furthermore, we need to consistently calculate $\frac{1}{N}$ corrections to all experimentally accessible quantities, such as masses, widths and cross-sections. After these quantities have been calculated and if -- importantly -- they are in acceptable agreement with experimental values, the presence or absence of the resonances predicted in section \ref{sec:resonances} will determine if our approach could be viable alternative to the celebrated Higgs mechanism.

We expect to report on progress on some of the above tasks in follow-up studies in the near future.

\section*{Acknowledgments}

This work was supported by the Department of Energy, DOE award No DE-SC0017905. PR would like to thank Scott Lawrence for helpful discussions and hospitality at Los Alamos National Laboratory during the early stages of this work. We also thank Sebastian Dawid, Seth Koren, Alex Soloviev, Gowri Sundaresan and Clifford Taubes for helpful comments on the manuscript.

\textbf{This work does not contain AI generated content or similar contaminants.}

\begin{appendix}
  \section{Including the Faddeev-Popov Determinant}
  \label{app1}

  In this appendix, we redo the calculation in the main text, but with the Faddeev-Popov explicitly included (not absorbed into a redefinition of the Higgs potential). The starting point is thus the path integral
  \be
  Z=\int {\cal D}\vec{\phi}\left|\vec{\phi}\right| {\cal D}A e^{-S_E}\,,
  \ee
  with $S_E$ given in (\ref{def}). We now introduce an auxiliary field $\sigma(x)$ as in Ref.~\cite{Romatschke:2023ztk}, so that
\be
  Z=\int {\cal D}\vec{\phi}{\cal D}\sigma \sqrt{\sigma}\delta(\sigma-\vec{\phi}^2) {\cal D}A e^{-S_E}\,,
  \ee
and where the Higgs potential in $S_E$ has been replaced using the delta function as
\be
\frac{\lambda_B}{N}\left(\vec{\phi}^2\right)^2\rightarrow \frac{\lambda_B}{N}\sigma^2\,.
\ee
Introducing a second auxiliary field $\zeta$ in order to exponentiate the $\delta$-function leads to
\be
  Z=\int {\cal D}\vec{\phi}{\cal D}\zeta {\cal D}A e^{-\int_x\left[\left(D_\mu \vec{\phi}\right)^\dagger D_\mu \vec{\phi}+\frac{1}{4}F_{\mu\nu}F_{\mu\nu}+i\zeta \vec{\phi}^2\right]}{\cal D}\sigma \sqrt{\sigma}e^{-\int_x\left[\frac{\lambda_B}{N}\sigma^2-i\zeta \sigma\right]}\,.
  \ee
Note that without the Faddeev-Popov determinant $\sqrt{\sigma}$, the path integral over $\sigma$ reproduces the result (\ref{Zthree3}) in the main text. With the determinant, it is useful to affect a change of coordinates $\sigma=\rho^2$ such that up to a constant factor
\be
\int {\cal D}\sigma \sqrt{\sigma}e^{-\int_x\left[\frac{\lambda_B}{N}\sigma^2-i\zeta \sigma\right]}=\frac{\cal D}{{\cal D} \zeta}\int {\cal D}\rho e^{-\int_x\left[\frac{\lambda_B}{N}\rho^4-i\zeta \rho^2\right]}\,.
\ee
The path integral over $\rho$ can be done in closed form as a product of modified Bessel functions $K_{\frac{1}{4}}$. However, in the continuum limit, the resulting expression simplifies and one gets up to a constant
\be
\int {\cal D}\sigma \sqrt{\sigma}e^{-\int_x\left[\frac{\lambda_B}{N}\sigma^2-i\zeta \sigma\right]}=\frac{\cal D}{{\cal D} \zeta} e^{-\int_x \frac{N \zeta^2}{8\lambda_B}}\,.
\ee
As a consequence, the partition function with the determinant included reads
\be
Z=\int {\cal D}\vec{\phi}{\cal D}\zeta \zeta {\cal D}A e^{-\int_x\left[\left(D_\mu \vec{\phi}\right)^\dagger D_\mu \vec{\phi}+\frac{1}{4}F_{\mu\nu}F_{\mu\nu}+i\zeta \vec{\phi}^2+\frac{N \zeta^2}{8\lambda_B}\right]}\,,
\ee
which should be compared to Eq.~(\ref{Zthree3}) in the main text. A further change of coordinates to
\be
u(x)=\zeta^2(x)\,,
\ee
then leads to
\be
Z=\int {\cal D}A {\cal D}u e^{-\frac{N}{2}{\rm tr}\ln\left[-\Box+i \sqrt{u}+\frac{e_B^2}{N}A_\mu A_\mu\right]-\int_x\left[\frac{N u}{8\lambda_B}+\frac{1}{4}F_{\mu\nu}F_{\mu\nu}\right]}\,,
\ee
where the fields $\vec{\phi}$ have been integrated out, cf. (\ref{flumS}). Performing an expansion in $\frac{1}{N}$ in the exponent then leads to the expanded action
\be
\tilde S_2=\frac{N}{2}{\rm tr}\ln\left[-\Box+i\sqrt{u}\right]+\frac{e_B^2}{2}{\rm tr}\left[\tilde \Delta A_\mu A_\mu\right]+\int dx\left[\frac{N u}{8\lambda_B}+\frac{1}{4}F_{\mu\nu}F_{\mu\nu}\right]\,,
\ee
with $\tilde \Delta=\left[-\Box+i \sqrt{u}\right]^{-1}$, cf. Eq.~(\ref{S2}). Expanding $u(x)$ around a global zero mode and fluctuations
\be
u(x)=u_0+\nu(x)\,,
\ee
we have
\be
\frac{N}{2}{\rm tr}\ln\left[-\Box+i \sqrt{u}\right]+\int_x \frac{N u}{8\lambda_B}=\frac{N}{2}{\rm tr}\ln\left[-\Box+i \sqrt{u_0}\right]+\int_x \frac{N u_0}{8\lambda_B}+\frac{N}{2} \int_{x,y} \nu D^{-1} \nu\,,
\ee
where the term linear in $\nu(x)$ vanishes. The saddle point condition for $u_0$ is
\be
 \tilde \Delta_0(0)=\frac{i\sqrt{u_0}}{2\lambda_B}\,,\quad \tilde \Delta_0(x)=\int \frac{d^4p}{(2\pi)^4} \frac{e^{i p\cdot x}}{p^2+i\sqrt{u_0}}\,,
\ee
which can be used to express the auxiliary field propagator as 
\be
D(k)=\frac{4 u_0}{\frac{1}{4\lambda_B}+\Pi(k)}\,,\quad \Pi(x)=\frac{1}{2}\tilde\Delta^2_0(x)\,.
\ee
Performing the same steps leading up to the effective potential (\ref{effpot}) in the main text, we find
\ba
     \label{effpotnew}
        \frac{{\cal V}}{{\rm vol}}&=&\frac{1}{2}\int \frac{d^4p}{(2\pi)^4}\ln\left[p^2+i\sqrt{u_0}\right]+\frac{u_0}{8\lambda_B}+\frac{1}{2N}\int \frac{d^4p}{(2\pi)^4}\ln \left[\frac{1}{4\lambda_B}+\Pi(p)\right]\nonumber\\
        &&+\frac{1}{2N}\int \frac{d^4p}{(2\pi)^4}\ln {\rm det}\left[\left(p^2+  i\sqrt{u_0} \frac{e_B^2}{2\lambda_B}\right) \delta_{\mu\nu}-p_\mu p_\nu\right]+{\cal O}(N^{-2})\,,
        \ea
        e.g. exactly the same as (\ref{effpot}) with the replacement $\lambda_B\rightarrow 2 \lambda_B$. Because of this, the calculation goes through as before. In particular we find the same leading order large N results for $\bar m_{\rm Higgs},\bar m_A$ in terms of $\Lambda_{\overline{\rm MS}}$ as in (\ref{LOhiggs}), (\ref{LOma}), but the running coupling changes from (\ref{LOrunco}) to
        \be
        \lambda_R(\bar\mu)=\frac{8\pi^2}{\ln \frac{\Lambda_{\overline{\rm MS}}^2}{\bar\mu^2}}+{\cal O}(N^{-1})\,,
        \ee
        which will change the phenomenology. To summarize, explicitly including the Faddeev-Popov determinant is possible in our framework, and does not lead to qualitative changes. 

\end{appendix}

\bibliography{EW}
\end{document}